\documentclass[letterpaper]{article} 
\usepackage{aaai2026}  
\usepackage{times}  
\usepackage{helvet}  
\usepackage{courier}  
\usepackage[hyphens]{url}  
\usepackage{graphicx} 
\usepackage{natbib}  
\usepackage{caption} 
\usepackage{subfigure}
\usepackage{multirow}
\usepackage{booktabs}
\usepackage[ruled,vlined,linesnumbered]{algorithm2e}
\usepackage{tikz} 
\usepackage{pgfplots} 
\pgfplotsset{compat=1.18} 
\usepackage{amsmath} 
\usepackage{tcolorbox}
\newtcolorbox{outbox}[1]{colback=blue!5!white,colframe=blue!75!black,fonttitle=\bfseries,title=#1}

\definecolor{5}{RGB}{156,209,255}
\definecolor{77}{RGB}{179,222,105}
\definecolor{1}{RGB}{141,211,199}
\definecolor{2}{RGB}{255,255,179}
\definecolor{3}{RGB}{190,186,218}
\definecolor{44}{RGB}{251,128,114}
\definecolor{66}{RGB}{253,180,98}
\definecolor{8}{RGB}{251,154,153}
\definecolor{99}{RGB}{31,120,180}
\definecolor{6}{RGB}{107,94,139}
\definecolor{9}{RGB}{78,129,121}
\definecolor{7}{RGB}{205,135,113}
\definecolor{4}{RGB}{127,66,82}
\definecolor{framcolor}{RGB}{176,196,222}
\newtcolorbox{mybox}[1]{colback=white, colframe=framcolor,coltitle=black,,fonttitle=\bfseries,title=#1}

\title{Diagnostic-Guided Dynamic Profile Optimization for LLM-based User Simulators in Sequential Recommendation}
\author {
    Hongyang Liu\textsuperscript{\rm 1},
    Zhu Sun\textsuperscript{\rm 2}\thanks{Corresponding author},
    Tianjun Wei\textsuperscript{\rm 3},
    Yan Wang\textsuperscript{\rm 1}\footnotemark[1],
    Jiajie Zhu\textsuperscript{\rm 1},
    Xinghua Qu\textsuperscript{\rm 4}
}
\affiliations {
    \textsuperscript{\rm 1}School of Computing, Macquarie University, Australia\\
    \textsuperscript{\rm 2}Information Systems Technology and Design, Singapore University of Technology and Design, Singapore\\
    \textsuperscript{\rm 3}College of Computing and Data Science, Nanyang Technological University, Singapore\\
    \textsuperscript{\rm 4}Bytedance Seed, Singapore\\
    hongyang.liu2@hdr.mq.edu.au, zhu\_sun@sutd.edu.sg, tjwei2-c@my.cityu.edu.hk\\ yan.wang@mq.edu.au, jiajie.zhu@mq.edu.au, xinghua.qu1@bytedance.com
}

\begin{document}

\maketitle

\begin{abstract}
Recent advances in large language models (LLMs) have enabled realistic user simulators for developing and evaluating recommender systems (RSs). However, existing LLM-based simulators for RSs face two major limitations: (1) static and single-step prompt-based inference that leads to inaccurate and incomplete user profile construction; (2) unrealistic and single-round recommendation-feedback interaction pattern that fails to capture real-world scenarios. To address these limitations, we propose DGDPO (\textbf{D}iagnostic-\textbf{G}uided \textbf{D}ynamic \textbf{P}rofile \textbf{O}ptimization), a novel framework that constructs user profile through a dynamic and iterative optimization process to enhance the simulation fidelity. 
Specifically, DGDPO incorporates two core modules within each optimization loop: firstly, a specialized LLM-based diagnostic module, calibrated through our novel training strategy, accurately identifies specific defects in the user profile. Subsequently, a generalized LLM-based treatment module analyzes the diagnosed defect and generates targeted suggestions to refine the profile. Furthermore, unlike existing LLM-based user simulators that are limited to single-round interactions, we are the first to integrate DGDPO with sequential recommenders, enabling a bidirectional evolution where user profiles and recommendation strategies adapt to each other over multi-round interactions. Extensive experiments conducted on three real-world datasets demonstrate the effectiveness of our proposed framework.
\end{abstract}


\section{Introduction}
User simulators~\cite{Balog:2024:FnTIR}, which aim to mimic user behaviors and decision-making patterns, are crucial for advancing recommender systems (RSs)~\cite{luo2022mindsim}. 
They offer a controllable and cost-effective paradigm for developing and evaluating RSs, reducing the reliance on real-user experiments that are often expensive and ethically sensitive. While early user simulators ranged from rule-based~\cite{rohde2018recogym} to deep learning-based approaches~\cite{shi2019virtual, zhao2021usersim}, these traditional methods fail to explicitly represent users' complex semantic preferences and generate diverse user interaction behaviors. To address these issues, Large Language Models (LLMs) have recently emerged as a new paradigm. With their advanced cognitive and generative capabilities, LLMs are typically integrated into an agent framework~\cite{park2023generative} that incorporates components such as Profile, Memory, and Action to emulate user interactions with RSs. For example, Agent4Rec~\cite{zhang2024generative} and RecAgent~\cite{wang2025user} prompt LLMs to generate user profile, enabling them to simulate diverse behaviors that mirror the complexity of real users.

However, existing LLM-based user simulators face two major limitations: \textbf{(1)} \textit{the inaccuracy and incompleteness of profile construction due to the static and single-step prompt-based inference}. By generating the user profile only at initialization, user simulators produce a static representation that cannot adapt to the evolving user interests across multi-round interactions.
Thus, any inaccuracy or incompleteness in this initial profile persists uncorrected, causing the simulated behavior to progressively diverge from real user actions throughout the simulation; \textbf{(2)} \textit{the unrealistic and single-round recommendation-feedback interaction pattern}. Most existing simulators are typically integrated with static recommenders, e.g., MF~\cite{koren2009matrix}, failing to simulate authentic, multi-round interactions where user profile and RS strategies should adapt and evolve based on the latest feedback.
\begin{figure}[t]
\raggedright 
\begin{tikzpicture}
\begin{axis}[
    width=0.58\columnwidth, 
    height=3.8cm,           
    ybar,
    enlarge x limits=0.25,
    bar width=8pt,
    ylabel={Accuracy (\%)},
    ylabel style={yshift=-5pt, font=\footnotesize},
    ymin=0, ymax=100,
    ymajorgrids=true,
    grid style={dashed, gray!50},
    symbolic x coords={ML-1M, Books, Movies \& TV},
    xtick=data,
    xticklabel style={font=\scriptsize, align=center}, 
    yticklabel style={font=\scriptsize},
    legend style={
        at={(1.02,0.6)}, 
        anchor=west,
        font=\footnotesize, 
        draw=none,
        legend cell align=left, 
        text width=3.5cm, 
        align=left,       
        row sep=20pt,
    },
    legend image post style={yshift=-2.5pt},
]
\addplot coordinates { (ML-1M, 63.5) (Books, 59.0) (Movies \& TV, 65.9) };
\addplot coordinates { (ML-1M, 94.1) (Books, 89.4) (Movies \& TV, 93.1) };

\legend{Prompt-based general-purpose LLM (GPT-4o-mini), Specialized LLM-based diagnostic module (Qwen3-0.6B)}
\end{axis}
\end{tikzpicture}
\captionsetup{skip=0.6pt}
\caption{Accuracy comparison for profile defect identification. Our specialized LLM-based diagnostic module significantly outperforms the prompt-based general-purpose LLM.}
\label{fig:diag_acc_v2}
\end{figure}
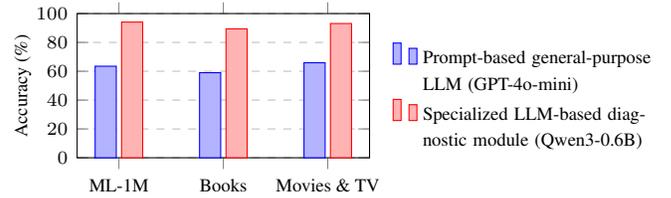

Therefore, we propose a novel \textbf{D}iagnostic-\textbf{G}uided \textbf{D}ynamic \textbf{P}rofile \textbf{O}ptimization (DGDPO) framework for a more realistic simulation. However, developing this framework faces two primary challenges. \textbf{(CH1)}: \textit{How to build a reliable mechanism for dynamic profile construction?} 
While existing methods~\cite{shu2024rah} like self-reflection~\cite{sun2024large} attempt iterative refinement, they often rely on general-purpose LLMs, thus may not be reliable due to potential hallucinations. As shown in Fig.~\ref{fig:diag_acc_v2}, directly prompting  general-purpose LLMs only achieves an average accuracy of 62.78\% on profile defect identification.
This highlights the necessity of a more targeted strategy that can accurately diagnose the specific type of defect, such as inaccurate and/or incomplete, and then apply a tailored refinement. 
To achieve this, we first design a specialized LLM-based diagnostic module through domain-adaptive pre-training and defect-specific fine-tuning, enabling reliable identification of profile defects. As shown in Fig.~\ref{fig:diag_acc_v2}, the module achieves an average accuracy of 92.20\%, significantly outperforming the general-purpose LLM.
We then design a generalized LLM-based treatment module that uses this precise diagnosis to analyze the defect and generate targeted refinements for the profile.
Instead of single-step prompting, our framework iteratively optimizes the profile by traversing the user's interaction history step-by-step, to dynamically enhance its accuracy and comprehensiveness.
\textbf{(CH2)}: \textit{How to achieve realistic multi-round interactions where user profiles and recommender strategies mutually influence and evolve over time?} Inspired by the multi-round interaction nature of real-world recommendation scenarios, we propose to incorporate the user simulator with sequential recommenders (SRs)~\cite{sun2025llm4rsr}. This establishes a genuine bidirectional evolution: the user profile is updated based on interactions,
while the SRs adaptively adjust their recommendation strategies in response to the evolving user behavior.

Our main contributions are three-fold: 
\textbf{(1)} To address \textbf{CH1}, we propose a novel framework \textbf{DGDPO}, which integrates a specialized LLM-based diagnostic module and a generalized LLM-based treatment module. By iteratively optimizing the user profile through step-by-step traversal of user interactions, the framework dynamically refines the user profile, resulting in a more accurate and comprehensive representation.
\textbf{(2)} To address \textbf{CH2}, we establish a realistic and multi-round interaction between the user simulator and sequential recommenders. This enables a bidirectional evolution of user profiles and recommendation strategies, offering a more credible evaluation environment. \textbf{(3)} We conduct extensive experiments on three real-world datasets, empirically validating the effectiveness of our DGDPO framework.

\section{Related Works}
\noindent\textbf{Traditional User Simulators for RSs.} Early user simulators for RSs progressed from rule-based approaches~\cite{rohde2018recogym, ie2019recsim, shi2019pyrecgym} to advanced data-driven techniques~\cite{huang2020keeping}, including methods based on reinforcement learning (RL)~\cite{shi2019virtual,zou2020pseudo}, generative adversarial networks (GANs)~\cite{chen2019generative, shi2019virtual, zhao2021usersim}, and transformers~\cite{zhao2023kuaisim,afzali2023usersimcrs}. However, they lack external context and knowledge, as well as the sophisticated cognitive reasoning mechanism that real users leverage during decision-making processes.

\noindent\textbf{LLM-based User Simulators for RSs.}
The advent of LLMs has introduced novel paradigms for user simulation in RSs~\cite{cai2025agentic,ma2025pub,wei2025mirroring}. LLM-based simulators have been widely adopted across various applications, including evaluating conversational systems~\cite{friedman2023leveraging,zhu2024reliable, zhu2025llm} and generating high-quality interaction data for RL-based algorithms~\cite{corecco2024llm,ebrat2024lusifer,zhang2025llm}. Current works on user profile construction can be categorized into two types. 
Some works~\cite{zhang2024generative, wang2025user} create user profile once at initialization, resulting in static representations that cannot capture evolving user preferences. Others employ self-reflection~\cite{zhang2024agentcf,shu2024rah} for iterative refinement.
However, they suffer from two limitations: (1) they adopt general-purpose LLMs which may generate unreliable profile due to potential hallucinations;
(2) they rely on unrealistic single-round interactions, lacking dynamic adaptation to user feedback and failing to simulate the bidirectional evolution between user profile and recommendation strategies in realistic interactions.

\noindent\textbf{Sequential Recommenders.}
SRs have evolved from early approaches combining Markov Chains with matrix factorization~\cite{rendle2010factorizing}, to deep learning models. These include CNNs for local pattern extraction~\cite{tang2018personalized}, RNNs to capture temporal dependencies~\cite{hidasi2015session, zhu2017next, zhu2019query}, attention mechanisms to focus on relevant historical items~\cite{li2017neural,kang2018self,liu2018stamp,sun2019bert4rec}, and GNNs to model high-order relationships within sequential data~\cite{wu2022gcrec,yu2024learning,li2024global,liu2024selfgnn}. Despite their success, they have primarily been evaluated in a static mode, which may fail to fully reflect the dynamic complexity of real-world user interactions.

\section{The Proposed Method}
\begin{figure*}[t]
\centering
\includegraphics[width=1\textwidth]{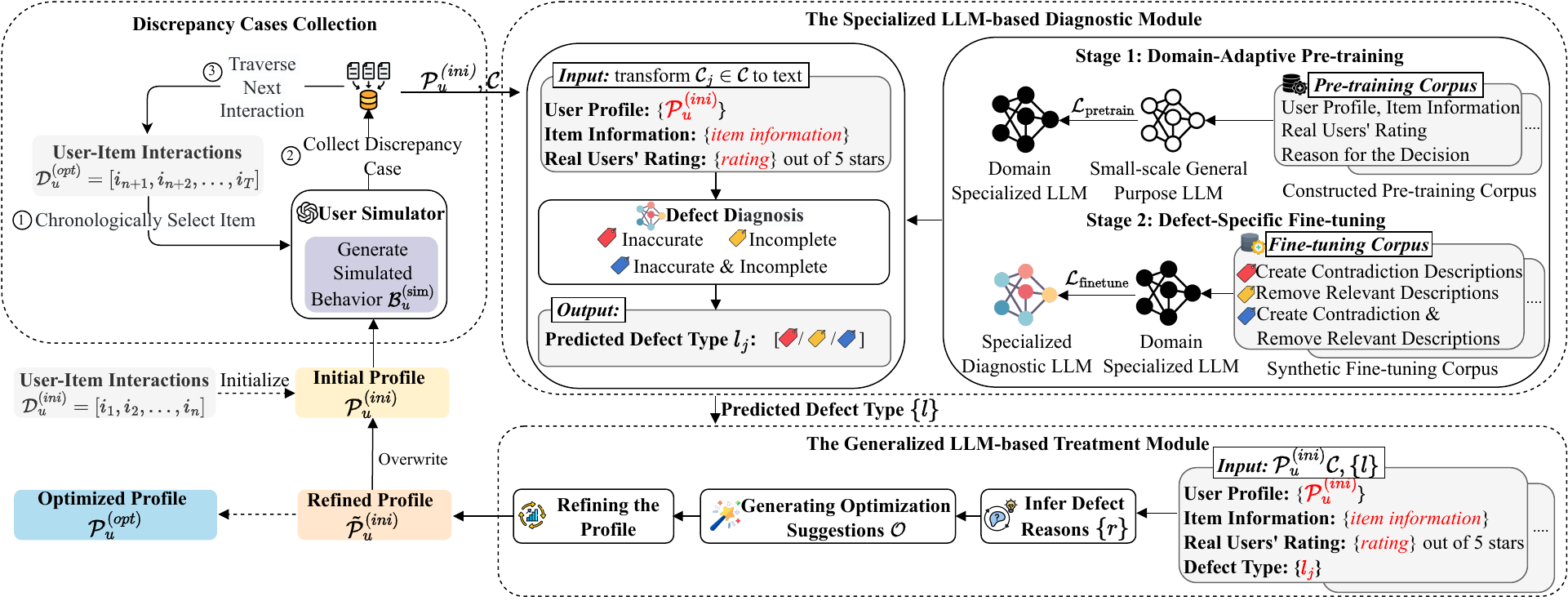}
\caption{The overall framework of our proposed DGDPO.}
\label{fig:framework_v3}
\end{figure*}

\noindent\textbf{Framework Overview.} 
DGDPO consists of two core modules that iteratively optimize the user profile in a dynamic manner, as shown in Fig.~\ref{fig:framework_v3}. First, a specialized LLM-based diagnostic module identifies profile defects by detecting discrepancies between simulated and real user behavior. Then, a generalized LLM-based treatment module generates targeted refinements to optimize the profile based on the diagnosis.
This optimization process is executed iteratively across a user's historical interaction sequence (i.e., step-by-step traversal of user interactions). For a more realistic simulation, DGDPO is further incorporated with SRs to establish multi-round interactions, enabling bidirectional evolution of user profiles and recommendation strategies.

\subsection{The DGDPO Framework (CH1)}
Given a user's historical interaction sequence $\mathcal{D}_u = [i_1, i_2, ..., i_T]$ ordered by timestamp, we split it into two portions: $\mathcal{D}_u^{(ini)} = [i_1, i_2,..., i_n]$, and $\mathcal{D}_u^{(opt)} = [i_{n+1},..., i_T]$. For effective and efficient user profile construction, we first create an initial user profile $\mathcal{P}_u^{(ini)}$ for each simulated user $u$. It is derived from $\mathcal{D}_u^{(ini)}$ via a single-step prompt-based inference method~\cite{zhang2024generative}, serving as the foundation user profile. Then, we leverage $\mathcal{D}_u^{(opt)}$ to dynamically optimize $\mathcal{P}_u^{(ini)}$ via our proposed DGDPO, ultimately obtaining an optimized user profile $\mathcal{P}_u^{(opt)}$. 

This optimization process is triggered by collecting discrepancy cases where the \textit{simulated behavior} $\mathcal{B}_u^{(\mathrm{sim})}$ of user $u$ mismatches his \textit{real behavior} $\mathcal{B}_u^{(\mathrm{real})}$ observed in the datasets (i.e., the simulator declines to interact with an item that the real user has interacted with). 
Specifically, we present the first item $i_{n+1} \in \mathcal{D}_u^{(opt)}$ to the user simulator built upon $\mathcal{P}_u^{(ini)}$. Then, the user simulator determines whether to interact (i.e., click) with $i_{n+1}$ and produces an explanatory reason for the decision. We traverse items in $\mathcal{D}_u^{(opt)}$ to collect a batch ($N_1$) of such discrepancy cases, which then serve as the input for our optimization workflow.

\subsubsection{The Specialized LLM-based Diagnostic Module.}
\label{sec: diagnostic_module}
Given a discrepancy case, this module aims to accurately identify the defect of the user profile via a specialized LLM. As such, it overcomes the inherent unreliability of general-purpose LLMs for such specific diagnostic tasks, providing more accurate signals to guide the optimization process.

First, we categorize each profile defect into one of three types based on its root cause: (1) \textit{Inaccurate}: the profile contains preference descriptions that contradict the real user's observed behavior. (2) \textit{Incomplete}: the profile lacks the specific preference descriptions required to explain the user's interaction with a given item. (3) \textit{Inaccurate \& Incomplete}: the profile exhibits both of the aforementioned issues. Then, we formally define this diagnostic task as a function $f_{\text{diag}}$. Given a user profile $\mathcal{P}_u$ and the corresponding discrepancy case $\mathcal{C}_j$, it outputs the corresponding defect type $l = f_{\text{diag}}(\mathcal{P}_u, \mathcal{C}_j)$ from the three defined types.

The key challenge now is how to accurately predict $l$. A straightforward way is to directly prompt general-purpose LLMs. However, their performance may be suboptimal due to limited domain knowledge and inherent hallucination (as illustrated in Fig.~\ref{fig:diag_acc_v2}). Therefore, we employ a delicately designed training strategy (i.e., domain-adaptive pre-training and defect-specific fine-tuning) to build a specialized LLM, transforming it into a specialized expert in profile defect identification by injecting extensive domain knowledge.

\textit{\textbf{Domain-Adaptive Pre-training.}}
This stage is to inject the model with the underlying semantics and reasoning patterns in recommendation. By pre-training on a large corpus of simulated interactions, the model learns the complex relationships between user profiles, item attributes, and interaction decisions, providing a robust foundation for the downstream defect identification task.

To construct the pre-training corpus, for each user $u$, we leverage his initial sequence data $\mathcal{D}_u^{(ini)}$. For each item $i_t \in \mathcal{D}_u^{(ini)}$, we instruct the user simulator with profile $\mathcal{P}_u^{(ini)}$ to make a binary interaction decision for $i_t$ and generate a textual rationale. To ensure the data quality, we only consider the case where the user simulator's decision matches with the real user $u$'s behavior (i.e., choosing to interact with $i_t$) and 
user $u$ gave item $i_t$ a high rating ($\geq$ 3 out of 5). Then, the user simulator's output, together with $\mathcal{P}_u^{(ini)}$ and relevant information (e.g., title, genres, and rating) of $i_t$, is used to construct a comprehensive corpus of structured texts.

Given the constructed corpus, we perform domain-adaptive pre-training on a small-scale general-purpose LLM using a language modeling objective, where the goal is to predict the next token in a sequence. This encourages the LLM to internalize the domain-specific logical flow of profile-based user simulation, defined as:
\begin{equation}
\mathcal{L}_{\text{pretrain}} = -\textstyle\sum_{{\scriptstyle t=1}}^{{\scriptstyle T_{Pre}}} \log P(x_t | x_{<t}, \theta),
\end{equation}
where $x = (x_1, \ldots, x_{T_{Pre}})$ represents a text sequence from the corpus with length $T_{Pre}$, and $\theta$ are the model parameters.

\textit{\textbf{Defect-Specific Fine-tuning.}} 
Next, we further fine-tune the LLM for the specific profile defect diagnosis task. A primary challenge lies in the absence of available, human-annotated data for defect identification. To overcome this, we develop a synthetic data generation pipeline that creates a high-quality, labeled corpus for fine-tuning purposes. 

The pipeline starts with the high-quality, non-defective profile-item pairs that constitute our pre-training corpus. Building upon these non-defective pairs, we subsequently perform a series of targeted modifications to generate synthetic user profiles with different defect types: (1) \textit{Inaccurate}: to create a contradiction, we first identify preference descriptions in the profile relevant to the target item (e.g., its genre). We then generate a new description with the opposite sentiment (e.g., transforming ``enjoys comedy films" to ``dislikes comedy films") and use it to replace the original statement. (2) \textit{Incomplete}: to simulate a lack of information, we randomly remove a proportion of descriptions relevant to the target item from the profile, making the profile insufficient to explain the user's interaction. (3) \textit{Inaccurate \& Incomplete}: we first apply the ``incomplete" modification, followed by the ``inaccurate" modification.
Using these synthetic profiles, we then format the data for instruction fine-tuning, where each sample has four parts: \textit{System}, \textit{Instruction}, \textit{Input}, and \textit{Output}.

To fine-tune the model, we employ a next token prediction loss objective with a modification: the loss is only calculated on the tokens in the \textit{Output} field. The tokens from the other fields are masked from the loss calculation, forcing the model to focus exclusively on more accurate profile defect diagnosis. Thus, the fine-tuning objective is to minimize the  loss only over the target response tokens:

\begin{equation}
\mathcal{L}_{\text{finetune}} = -\sum\nolimits_{t \in T_{\text{Output}}} \log P(x_t | x_{<t}, \theta),
\end{equation}
where $T_{\text{Output}}$ is the set of token indices corresponding to the \textit{Output} field, and $\theta$ are the model parameters.

In summary, our delicately designed training strategy transforms the general-purpose LLM into a highly accurate specialized diagnostic module. It provides a trustworthy and structured diagnostic signal that is essential for guiding the subsequent treatment module.

\newcommand{\Init}{\text{Init}}
\newcommand{\Simu}{\text{Simu}}
\newcommand{\Real}{\text{Real}}
\newcommand{\DiagnoseDefect}{\text{DiagnoseDefect}}
\newcommand{\GenOpt}{\text{GenOpt}}
\newcommand{\Refine}{\text{Refine}}
\newcommand{\Infer}{\text{Infer}}
\begin{algorithm}[t]
\small
\SetCommentSty{small}
\caption{\textsc{DGDPO}}\label{alg:iterative-optimization-v2} 
\LinesNumbered
\KwIn{\(\mathcal{D}_u^{(ini)}\), \(\mathcal{D}_u^{(opt)}\), \(N_1\)}
\KwOut{\(\mathcal{P}_u^{(opt)}\)}
    \(\mathcal{C}\) \(\leftarrow\) [ ]\tcp*{Initialize}
    \(\mathcal{P}_u^{(ini)}\) \(\leftarrow\) Init(\(\mathcal{D}_u^{(ini)}\))\tcp*{Inference}
    \For{item $i$ in \(\mathcal{D}_u^{(opt)}\)}{
        \(\mathcal{B}_u^{(\mathrm{sim})}\) \(\leftarrow\) Simu(\(\mathcal{P}_u^{(ini)}\), $i$)\tcp*{Get simulated behavior}
        \(\mathcal{B}_u^{(\mathrm{real})}\) \(\leftarrow\) Real(\(\mathcal{D}_u^{(opt)}\), $i$)\tcp*{Get real behavior}
        \If{\(\mathcal{B}_u^{(\mathrm{sim})}\) \(\neq\) \(\mathcal{B}_u^{(\mathrm{real})}\)}{
            $\mathcal{C}.append(i, \mathcal{B}_u^{(\mathrm{sim})}$, \(\mathcal{B}_u^{(\mathrm{real})}\))\;
            \If{\(|\mathcal{C}| = N_1\)}{
                \tcp{The Diagnostic Module}
                $\{l\} \gets f_{\text{diag}}(\mathcal{P}_u^{(ini)}, \mathcal{C})$\;
                \tcp{The Treatment Module}
                $\{r\} \gets \Infer(\mathcal{P}_u^{(ini)}, \mathcal{C}, \{l\})$
           
                $\mathcal{O} \gets \GenOpt(\mathcal{P}_u^{(ini)}, \mathcal{C}, \{l\}, \{r\})$\;
                $\tilde{\mathcal{P}}_u^{(ini)} \gets \Refine(\mathcal{P}_u^{(ini)}, \mathcal{O})$
                \(\mathcal{C}\) \(\leftarrow\) [ ]\;
                \(\mathcal{P}_u^{(ini)}\) \(\leftarrow\) $\tilde{\mathcal{P}}_u^{(ini)}$;
            }
        }
    }
    \(\mathcal{P}_u^{(opt)}\) \(\leftarrow\) $\tilde{\mathcal{P}}_u^{(ini)}$\;
    \Return{${\mathcal{P}}_u^{(opt)}$}\;
\end{algorithm}

\subsubsection{The Generalized LLM-based Treatment Module.} 
This module is responsible for the complex reasoning and modification tasks. While the specialized diagnostic module excels at identifying what defects exist within a profile, understanding why these defects occur and determining how to address them effectively requires sophisticated common-sense reasoning and high-fidelity generative capabilities. Consequently, we employ a powerful general-purpose LLM, leveraging its superior analytical reasoning and text generation capabilities to carry out the following treatment process.

Specifically, we first prompt the LLM by providing it with the current user profile $\mathcal{P}_u^{(ini)}$, the target item $i$, and defect label $l$, to generate a comprehensive textual explanation (i.e., inferred reason $r$) towards the diagnosis made by the diagnosis module. Given the inferred reason, we further instruct the LLM to generate a set of concrete optimization suggestions (i.e., targeted modifications) denoted as $\mathcal{O}$. For instance, an \textit{inaccurate} defect may result in a suggestion to correct contradictory preferences, while an \textit{incomplete} defect would lead to a suggestion to add missing details.
Finally, we prompt the LLM to apply the suggestions $\mathcal{O}$ to produce an updated user profile $\tilde{\mathcal{P}}_u^{(ini)}$. In this step, specific constraints are enforced to (1) maintain the profile's overall coherence and descriptive style, and (2) avoid mentioning specific item details, ensuring the resulting profile captures general user preferences.

\subsubsection{Iterative Optimization.}  
Given the updated profile $\tilde{\mathcal{P}}_u^{(ini)}$, our framework continues to traverse the remaining interactions in $\mathcal{D}_u^{(opt)}$ to collect the next batch of discrepancy cases. Once a batch is accumulated, the complete workflow is executed: the diagnostic module provides a diagnosis for each case, which subsequently guides the treatment module to produce the next updated profile. This process repeats until all interactions in $\mathcal{D}_u^{(opt)}$ are processed, ultimately yielding the final optimized user profile $\mathcal{P}_u^{(opt)}$. The complete process is detailed in Algorithm~\ref{alg:iterative-optimization-v2}.

\subsection{Bidirectional Evolution between User Profiles and Recommendation Strategies (CH2)}
Existing LLM-based user simulators are primarily integrated with static recommenders in page-by-page scenarios~\cite{zhang2024generative}, which essentially represent a one-time recommendation result split into multiple pages. 
This fails to capture realistic scenarios where users continuously interact with the RSs, leading to the dynamic evolution of both user profiles and recommendation strategies through user feedback and interactions. 
To enable realistic multi-round interactions, we propose to incorporate our optimized user simulator DGDPO with sequential recommenders (SRs). Unlike traditional static models, SRs excel at capturing temporal dynamics, which makes them ideal for modeling the bidirectional evolution of user preferences and recommendation strategies over time.

Specifically, in the multi-round interaction scenario, our DGDPO and SRs engage in a continuous interaction process that consists of four steps: (1) Based on the user's interaction history, the SRs generate a ranked list of candidate items $\mathcal{X}$ for the optimized simulator with profile \(\mathcal{P}_u^{(opt)}\), where one item is the ground truth, and the others are randomly sampled negative items; (2) The user simulator decides whether to interact with the recommended items, where at most one item can be selected for interaction in each round; (3) When an item is selected, the simulator performs an update to \(\mathcal{P}_u^{(opt)}\) to reflect the recent interaction pattern. In particular, we design two strategies for profile update: (i) \textit{Update without Ground Truth (w/o GT)}: the profile is updated using the item that the simulator chooses to interact with.  (ii) \textit{Update with Ground Truth (w/ GT)}: the profile is updated using the ground truth positive item to maintain alignment with real user behavior. In both strategies, the profile update is performed via a carefully controlled prompt that instructs the LLM to make minimal, incremental adjustments while preserving the profile's overall coherence; and (4) Based on this newly observed interaction, the SRs update their internal hidden state for next recommendation. This loop continues until the simulator either chooses to
exit or reaches the maximum rounds of interactions.

\section{Experiments and Results}
We conduct extensive experiments to answer five research questions\footnote{Our code and datasets are available at \url{https://github.com/hyllll/DGDPO}}. (\textbf{RQ1}) Can DGDPO improve the fidelity of user simulators? (\textbf{RQ2}) How do different profile update strategies and SRs impact the simulation fidelity in multi-round interaction scenarios? (\textbf{RQ3}) How do various components of DGDPO affect its performance? (\textbf{RQ4}) How do essential hyper-parameters affect DGDPO? (\textbf{RQ5}) How does DGDPO construct user profile with its optimization process?

\begin{table}[t]
\footnotesize
\setlength{\tabcolsep}{3.5pt}
\begin{tabular}{l|ccc}
\toprule
& ML-1M & Books & Movies \& TV \\\midrule
\#Users & 6,040 & 208,864 & 23,969 \\
\#Items & 3,416 & 241,725 & 25,830 \\
\#Interactions & 999,611 & 10,865,527 & 1,053,194 \\
Avg. Sequence Length & 198.41 & 49.16 & 48.59 \\
Sparsity & 95.16\% & 99.98\% & 99.83\% \\
\bottomrule
\end{tabular}
\caption{Dataset statistics after preprocessing.}
\label{tab:statistics}
\end{table}

\begin{table*}[!ht]
\footnotesize
\centering
\setlength{\tabcolsep}{2pt} 
\begin{tabular}{cl|cccc|cccc|cccc}
\toprule
\multicolumn{2}{c|}{\textbf{Datasets}} & \multicolumn{4}{c|}{\textbf{ML-1M}} & \multicolumn{4}{c|}{\textbf{Books}} & \multicolumn{4}{c}{\textbf{Movies \& TV}} \\ \midrule
\multicolumn{2}{c|}{Methods} & \multicolumn{1}{c}{Prec} & Recall & Acc & F1 & \multicolumn{1}{c}{Prec} & Recall & Accuracy & F1 & \multicolumn{1}{c}{Prec} & Recall & Acc & F1 \\ \midrule
\multicolumn{1}{c|}{Traditional} & KuaiSim
&0.1067 &0.6304 &0.6651 &0.1825          
&0.1423 &0.6107 &0.6728 &0.2308          
&0.0832 &0.5917 &0.6228 &0.1459   \\ \midrule
\multicolumn{1}{c|}{\multirow{4}{*}{LLM-based}} & Agent4Rec                    
&0.1031 &0.6250 &0.6600 &0.1770          
&0.1521 &0.6431 &0.6920  &0.2460		       		
&0.0823 &0.6013 &0.6322 &0.1448    \\ 		
\multicolumn{1}{c|}{} & LLM-US                       
&0.0978 &0.6208 &0.6702*  &0.1690 		      
&0.1457 &0.6243 &0.6832  &0.2363          		
&0.0765 &0.5984 &0.6281  &0.1357    \\ 		
\multicolumn{1}{c|}{} & RAH                          
&0.1104 &0.6413 &0.6557 &0.1884          		
&0.1565 &0.6348 &0.6874 &0.2511      		
&0.0854 &0.6078 &0.6359 &0.1498     \\ 		
\multicolumn{1}{c|}{} & AgentCF                      
&0.1157* &0.6424* &0.6652 &0.1961*	   
&0.1598* &0.6581* &0.6953* &0.2572*          		
&0.0871* &0.6105* &0.6394* &0.1524*     \\ 
\multicolumn{1}{c|}{} & DGDPO                      
&\textbf{0.1430} &\textbf{0.7365} &\textbf{0.7174} &\textbf{0.2395}	   		
&\textbf{0.2058} &\textbf{0.7556} &\textbf{0.7570} &\textbf{0.3235}         		 		
&\textbf{0.1038} &\textbf{0.6913} &\textbf{0.6884} &\textbf{0.1805}     \\ \midrule 		
\multicolumn{2}{c|}{\textit{Improve}}                                              
&23.60\% &14.65\% &7.04\% &22.13\%   
&28.78\% &14.81\% &8.87\% &25.82\%          
&19.17\% &13.24\% &7.66\% &18.44\%      \\ 
\bottomrule
\end{tabular}
\caption{Performance comparison on all datasets, where the best and runner-up results are highlighted in bold and marked by `*'; and `\textit{Improve}' indicates the relative improvements comparing the best and runner-up results. Statistical significance of the improvement is determined by a paired t-test with $p$-value $<$ 0.01.}
\label{tab:comparison}
\end{table*}

\pgfplotsset{
compat=1.11,
legend image code/.code={
\draw[mark repeat=2,mark phase=2]
plot coordinates {
(0cm,0cm)
(0.15cm,0cm)        
(0.3cm,0cm)         
};%
}
}
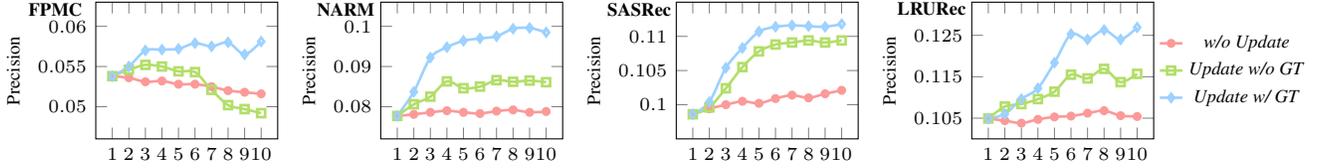
\begin{figure*}[t]
 \centering
 \subfigure{
  \begin{tikzpicture}
  \begin{axis}[
  width=4.0cm,  
  height=3.4cm, 
  ylabel={Precision},
  xlabel={},
  xmin=0, xmax=11,
  ymin=0.046, ymax=0.063,
  xtick={1,2,3,4,5,6,7,8,9,10},
  yticklabel style={/pgf/number format/.cd,fixed,precision=3},
  ytick={0.050, 0.055, 0.060},
  ylabel style = {font=\scriptsize},
  xlabel style = {font=\scriptsize},
  tick label style={font=\scriptsize},
  scaled ticks=false,
  title ={\textbf{FPMC}},
  title style={at={(-0.42,0.95)}, anchor=north west, font=\scriptsize},
  ]
 \addplot[color=8,
  solid,
  mark=*,
  mark options={solid},
  line width=1pt,
  mark size=1.2pt,								
  smooth] coordinates { 
            (1, 0.0538)
            (2, 0.0536)
            (3, 0.0531)
            (4, 0.0532)
            (5, 0.0528)
            (6, 0.0528)
            (7, 0.0525)
            (8, 0.052)
            (9, 0.0518)
            (10, 0.0516)
  };
 \addplot[ color=77, 
  solid,
  mark=square,
  mark options={solid},
  line width=1pt,mark size=1.5pt,									
  smooth] coordinates { 		
            (1, 0.0538)
            (2, 0.0546)
            (3, 0.0552)
            (4, 0.055)
            (5, 0.0544)
            (6, 0.0543)
            (7, 0.0521)
            (8, 0.0502)
            (9, 0.0497)
            (10, 0.0492)
   };
 \addplot[ color=5, 
  solid,
  mark=diamond,
  mark options={solid},
  line width=1pt,mark size=1.5pt,
  smooth] coordinates { 
            (1, 0.0538)
            (2, 0.055)
            (3, 0.057)
            (4, 0.0571)
            (5, 0.0572)
            (6, 0.0579)
            (7, 0.0575)
            (8, 0.058)
            (9, 0.0565)
            (10, 0.0581)
        };
 
 \end{axis}
 \end{tikzpicture}}
\subfigure{
  \begin{tikzpicture}
  \begin{axis}[
  width=4.0cm,
  height=3.4cm,
  ylabel={Precision},
  xlabel={},
  xmin=0, xmax=11,
  ymin=0.072, ymax=0.106,
  xtick={1,2,3,4,5,6,7,8,9,10},
  yticklabel style={/pgf/number format/.cd,fixed,precision=3},
  ytick={0.080, 0.090, 0.100},
  ylabel style = {font=\scriptsize},
  xlabel style = {font=\scriptsize},
  tick label style={font=\scriptsize},
  scaled ticks=false,
  title ={\textbf{NARM}},
  title style={at={(-0.40,0.95)}, anchor=north west, font=\scriptsize},
  ]
 \addplot[color=8,
  solid,
  mark=*,
  mark options={solid},
  line width=1pt,
  mark size=1.2pt,
  smooth] coordinates { 			
            (1, 0.0777)
            (2, 0.0781)
            (3, 0.0786)
            (4, 0.079)
            (5, 0.0786)
            (6, 0.0783)
            (7, 0.0789)
            (8, 0.0792)
            (9, 0.0786)
            (10, 0.0788)
  };
 \addplot[ color=77, 
  solid,
  mark=square,
  mark options={solid},
  line width=1pt,mark size=1.5pt,
  smooth] coordinates { 				
            (1, 0.0777)
            (2, 0.0806)
            (3, 0.0825)
            (4, 0.0863)
            (5, 0.0846)
            (6, 0.085)
            (7, 0.0866)
            (8, 0.0862)
            (9, 0.0865)
            (10, 0.0861)
   };
 \addplot[ color=5, 
  solid,
  mark=diamond,
  mark options={solid},
  line width=1pt,mark size=1.5pt,
  smooth] coordinates { 
            (1, 0.0777)
            (2, 0.0837)
            (3, 0.0922)
            (4, 0.0948)
            (5, 0.0964)
            (6, 0.097)
            (7, 0.0975)
            (8, 0.0994)
            (9, 0.0996)
            (10, 0.0985)
        };
 \end{axis}
 \end{tikzpicture}}
 \subfigure{
  \begin{tikzpicture}
  \begin{axis}[
  width=4.0cm,
  height=3.4cm,
  ylabel={Precision},
  xlabel={},
  xmin=0, xmax=11,
  ymin=0.095, ymax=0.115,
  xtick={1,2,3,4,5,6,7,8,9,10},
  yticklabel style={/pgf/number format/.cd,fixed,precision=3},
  ytick={0.100, 0.105, 0.110},
  ylabel style = {font=\scriptsize},
  xlabel style = {font=\scriptsize},
  tick label style={font=\scriptsize},
  scaled ticks=false,
  title ={\textbf{SASRec}},
  title style={at={(-0.43,0.95)}, anchor=north west, font=\scriptsize},
  ]
 \addplot[color=8,
  solid,
  mark=*,
  mark options={solid},
  line width=1pt,
  mark size=1.2pt,
  smooth] coordinates { 					
            (1, 0.0986)
            (2, 0.0994)
            (3, 0.1)
            (4, 0.1005)
            (5, 0.1002)
            (6, 0.1009)
            (7, 0.1014)
            (8, 0.101)
            (9, 0.1016)
            (10, 0.1021)
  };
 \addplot[ color=77, 
  solid,
  mark=square,
  mark options={solid},
  line width=1pt,mark size=1.5pt,
  smooth] coordinates { 	
            (1, 0.0986)
            (2, 0.0996)
            (3, 0.1024)
            (4, 0.1055)
            (5, 0.1078)
            (6, 0.1088)
            (7, 0.1091)
            (8, 0.1094)
            (9, 0.1091)
            (10, 0.1094)
   };
 \addplot[ color=5, 
  solid,
  mark=diamond,
  mark options={solid},
  line width=1pt,mark size=1.5pt,
  smooth] coordinates { 		
            (1, 0.0986)
            (2, 0.1004)
            (3, 0.1054)
            (4, 0.1083)
            (5, 0.1107)
            (6, 0.1114)
            (7, 0.1116)
            (8, 0.1115)
            (9, 0.1114)
            (10, 0.1118)
        };
 \end{axis}
 \end{tikzpicture}}
 \subfigure{
  \begin{tikzpicture}
  \begin{axis}[
  width=4.0cm,
  height=3.4cm,
  ylabel={Precision},
  xlabel={},
  xmin=0, xmax=11,
  ymin=0.100, ymax=0.133,
  xtick={1,2,3,4,5,6,7,8,9,10},
  yticklabel style={/pgf/number format/.cd,fixed,precision=3},
  ytick={(0.105, 0.115, 0.125)},
  ylabel style = {font=\scriptsize},
  xlabel style = {font=\scriptsize},
  tick label style={font=\scriptsize},
  scaled ticks=false,
  legend style={
  at={(0.98,0.5)}, 
  anchor=west,
  legend columns=1,
  font=\scriptsize,
  draw=none, 
  fill=none},
  title ={\textbf{LRURec}},
  title style={at={(-0.46,0.95)}, anchor=north west, font=\scriptsize},
  ]
 \addplot[color=8,
  solid,
  mark=*,
  mark options={solid},
  line width=1pt,
  mark size=1.2pt,
  smooth] coordinates { 
            (1, 0.1049)
            (2, 0.1044)
            (3, 0.1038)
            (4, 0.1047)
            (5, 0.1053)
            (6, 0.1055)
            (7, 0.1062)
            (8, 0.1068)
            (9, 0.1056)
            (10, 0.1054)
  };
 \addplot[ color=77, 
  solid,
  mark=square,
  mark options={solid},
  line width=1pt,mark size=1.5pt,
  smooth] coordinates {
            (1, 0.1049)
            (2, 0.1078)
            (3, 0.1084)
            (4, 0.1096)
            (5, 0.1114)
            (6, 0.1155)
            (7, 0.1146)
            (8, 0.1169)
            (9, 0.1137)
            (10, 0.1157)
   };
 \addplot[ color=5, 
  solid,
  mark=diamond,
  mark options={solid},
  line width=1pt,mark size=1.5pt,
  smooth] coordinates { 				
            (1, 0.1049)
            (2, 0.106)
            (3, 0.1096)
            (4, 0.1122)
            (5, 0.1184)
            (6, 0.1253)
            (7, 0.124)
            (8, 0.1263)
            (9, 0.124)
            (10, 0.1269)
        };
 \legend{\textit{w/o Update}, \textit{Update w/o GT}, \textit{Update w/ GT}}
 \end{axis}
 \end{tikzpicture}}
 \captionsetup{justification=centering}
 \caption{The impact of profile update strategies on DGDPO in multi-round interactions on Books dataset.}
\label{fig:interaction_analysis_book}
\end{figure*}

\subsection{Experimental Setup}
\subsubsection{Datasets.} 
We use three datasets: ML-1M~\cite{harper2015movielens}, Amazon Books, and Amazon Movies \& TV~\cite{ni2019justifying}. We filter users and items with less than 5 interactions (ML-1M) or less than 20 (Amazon datasets). Interactions are sorted chronologically, with max sequence lengths set to 200 for ML-1M and 50 for Amazon datasets. For each user, the last 10 items are for testing, the 11th most recent item is for validation, and the rest for training (i.e., \(\mathcal{D}_u\)). \(\mathcal{D}_u\) is further split into \(\mathcal{D}_u^{(ini)}\) and \(\mathcal{D}_u^{(opt)}\) by an \(\alpha:(1-\alpha)\) ratio.


\subsubsection{Baselines.} 
We compare DGDPO with five baselines for user profile construction in two categories. 
In traditional methods, \textbf{KuaiSim}~\cite{zhao2023kuaisim} is a transformer-based model for user profile construction. 
In LLM-based methods, \textbf{Agent4Rec}~\cite{zhang2024generative} utilizes a single-step prompt-based inference. \textbf{LLM-US}~\cite{zhang2025llm} generates explanations of user ratings as profile. \textbf{RAH}~\cite{shu2024rah} and \textbf{AgentCF}~\cite{zhang2024agentcf} employ self-reflection and collaborative reflection, respectively.
\textit{For multi-round interactions, we integrate the user simulator with four SRs}: 
\textbf{FPMC}~\cite{rendle2010factorizing}, \textbf{NARM}~\cite{li2017neural}, \textbf{SASRec}~\cite{kang2018self} and \textbf{LRURec}~\cite{yue2024linear}.

\subsubsection{Implementation Details.} Following state-of-the-arts~\cite{shu2024rah,zhang2024generative}, we randomly sample 1000 users from each dataset to serve as simulated users. Our diagnostic module is based on lightweight Qwen3-0.6B~\cite{yang2025qwen3}, while the treatment module employs gpt-4o-mini via its API with a temperature of 0. All LLM-based baselines also utilize GPT-4o-mini as their backbone. For DGDPO, we set $N_1=4$, $\alpha=0.6$, and  $\mathcal{X} = 20$ for all datasets. We build upon Agent4Rec~\cite{zhang2024generative} as the foundation user simulator and conduct ten-round interactions for each user. For KuaiSim and SRs methods, we tune them by Optuna (optuna.org) with 50 trials and train for up to 100 epochs with an early stopping mechanism~\cite{sun2022daisyrec}. All experiments are conducted on a V100 GPU.

\subsubsection{Evaluation Metrics.}
Following~\cite{luo2022mindsim,zhang2024generative}, we adopt Precision (Prec), Recall, Accuracy (Acc), and F1 Score (F1) as the evaluation metrics to measure the consistency between the simulated and real users. 

\subsection{Results and Analysis}
\subsubsection{Overall Comparison (RQ1).}
Table~\ref{tab:comparison} presents the performance comparison of all methods. 
The results show that DGDPO achieves the best performance across all datasets and metrics, with average relative improvements of 16.87\%, 18.82\%, and 14.63\% over the runner-up methods across the three datasets, respectively. These substantial improvements demonstrate the effectiveness of our method. In addition, we observe that DGDPO provides a particularly substantial boost in Precision, with relative improvements of 23.60\%, 28.78\%, and 19.17\% across the three datasets.
This indicates that DGDPO can accurately identify and guide the optimization of defective user profiles, generating more precise and faithful user representations that prevent incorrect item selections and significantly enhance simulation precision.

\pgfplotsset{
compat=1.11,
legend image code/.code={
\draw[mark repeat=2,mark phase=2]
plot coordinates {
(0cm,0cm)
(0.15cm,0cm)        
(0.3cm,0cm)         
};%
}
}
\begin{figure*}[t]
 \centering
 \subfigure{
  \begin{tikzpicture}
  \begin{axis}[
  width=4.0cm,
  height=3.4cm,
  ylabel={Precision},
  xlabel={},
  xmin=0, xmax=11,
  ymin=0.0482, ymax=0.05650,
  xtick={1,2,3,4,5,6,7,8,9,10},
  yticklabel style={/pgf/number format/.cd,fixed,precision=3},
  ytick={0.0490,0.0510,0.0530, 0.0550},
  ylabel style = {font=\scriptsize},
  xlabel style = {font=\scriptsize},
  tick label style={font=\scriptsize},
  scaled ticks=false,
  legend style={at={(0.35,0.99)}, font=\scriptsize, anchor=north,legend columns=1, draw=none, fill=none,},
  title ={\textbf{FPMC}},
  title style={at={(-0.42,0.99)}, anchor=north west, font=\scriptsize},
  ]
 \addplot[color=8,
  solid,
  mark=*,
  mark options={solid},
  line width=1pt,
  mark size=1.2pt,								
  smooth] coordinates { 
            (1, 0.0524)
            (2, 0.0521)
            (3, 0.0518)
            (4, 0.052)
            (5, 0.0515)
            (6, 0.0517)
            (7, 0.0513)
            (8, 0.0511)
            (9, 0.0514)
            (10, 0.051)
  };
 \addplot[ color=77, 
  solid,
  mark=square,
  mark options={solid},
  line width=1pt,mark size=1.5pt,									
  smooth] coordinates { 					
            (1, 0.0538)
            (2, 0.0534)
            (3, 0.053)
            (4, 0.0532)
            (5, 0.0526)
            (6, 0.0528)
            (7, 0.0525)
            (8, 0.052)
            (9, 0.0523)
            (10, 0.0516)	
   };
 \addplot[ color=5, 
  solid,
  mark=diamond,
  mark options={solid},
  line width=1pt,mark size=1.5pt,
  smooth] coordinates { 				
            (1, 0.0515)
            (2, 0.0512)
            (3, 0.051)
            (4, 0.0508)
            (5, 0.0504)
            (6, 0.0507)
            (7, 0.0505)
            (8, 0.0501)
            (9, 0.0503)
            (10, 0.0499)  
            };
 \end{axis}
 \end{tikzpicture}}
\subfigure{
  \begin{tikzpicture}
  \begin{axis}[
  width=4.0cm,
  height=3.4cm,
  ylabel={Precision},
  xlabel={},
  xmin=0, xmax=11,
  ymin=0.0690, ymax=0.0840,
  xtick={1,2,3,4,5,6,7,8,9,10},
  yticklabel style={/pgf/number format/.cd,fixed,precision=4},
  ytick={0.0700,0.0750,0.0800},
  ylabel style = {font=\scriptsize},
  xlabel style = {font=\scriptsize},
  tick label style={font=\scriptsize},
  scaled ticks=false,
  legend style={at={(0.63,0.52)}, font=\scriptsize, anchor=north,legend columns=1, draw=none, fill=none,},
  title ={\textbf{NARM}},
  title style={at={(-0.40,0.95)}, anchor=north west, font=\scriptsize},
  ]
 \addplot[color=8,
  solid,
  mark=*,
  mark options={solid},
  line width=1pt,
  mark size=1.2pt,
  smooth] coordinates { 			
            (1, 0.0724)
            (2, 0.0727)
            (3, 0.073)
            (4, 0.0733)
            (5, 0.073)
            (6, 0.0728)
            (7, 0.0732)
            (8, 0.0734)
            (9, 0.073)
            (10, 0.0731)
  };
 \addplot[ color=77, 
  solid,
  mark=square,
  mark options={solid},
  line width=1pt,mark size=1.5pt,
  smooth] coordinates { 						
            (1, 0.0777)
            (2, 0.0781)
            (3, 0.0786)
            (4, 0.079)
            (5, 0.0786)
            (6, 0.0783)
            (7, 0.0789)
            (8, 0.0792)
            (9, 0.0786)
            (10, 0.0788)
   };
 \addplot[ color=5, 
  solid,
  mark=diamond,
  mark options={solid},
  line width=1pt,mark size=1.5pt,
  smooth] coordinates { 					
            (1, 0.0718)
            (2, 0.072)
            (3, 0.0722)
            (4, 0.0725)
            (5, 0.0721)
            (6, 0.072)
            (7, 0.0723)
            (8, 0.0724)
            (9, 0.072)
            (10, 0.0721)
        };
 \end{axis}
 \end{tikzpicture}}
 \subfigure{
  \begin{tikzpicture}
  \begin{axis}[
  width=4.0cm,
  height=3.4cm,
  ylabel={Precision},
  xlabel={},
  xmin=0, xmax=11,
  ymin=0.0855, ymax=0.1050,
  xtick={1,2,3,4,5,6,7,8,9,10},
  yticklabel style={/pgf/number format/.cd,fixed,precision=3},
  ytick={0.0850,0.0900,0.0950,0.1000},
  ylabel style = {font=\scriptsize},
  xlabel style = {font=\scriptsize},
  tick label style={font=\scriptsize},
  scaled ticks=false,
  legend style={at={(0.60,0.52)}, font=\scriptsize, anchor=north,legend columns=1, draw=none, fill=none,},
  title ={\textbf{SASRec}},
  title style={at={(-0.43,0.95)}, anchor=north west, font=\scriptsize},
  ]
 \addplot[color=8,
  solid,
  mark=*,
  mark options={solid},
  line width=1pt,
  mark size=1.2pt,
  smooth] coordinates { 							
            (1, 0.0898)
            (2, 0.0904)
            (3, 0.0907)
            (4, 0.091)
            (5, 0.0908)
            (6, 0.0913)
            (7, 0.0916)
            (8, 0.0914)
            (9, 0.0918)
            (10, 0.0921)
  };
 \addplot[ color=77, 
  solid,
  mark=square,
  mark options={solid},
  line width=1pt,mark size=1.5pt,
  smooth] coordinates { 				
            (1, 0.0986)
            (2, 0.0994)
            (3, 0.1)
            (4, 0.1005)
            (5, 0.1002)
            (6, 0.1009)
            (7, 0.1014)
            (8, 0.101)
            (9, 0.1016)
            (10, 0.1021)
   };
 \addplot[ color=5, 
  solid,
  mark=diamond,
  mark options={solid},
  line width=1pt,mark size=1.5pt,
  smooth] coordinates { 				
            (1, 0.0885)
            (2, 0.089)
            (3, 0.0893)
            (4, 0.0896)
            (5, 0.0892)
            (6, 0.0897)
            (7, 0.09)
            (8, 0.0898)
            (9, 0.0902)
            (10, 0.0905)
        };
 \end{axis}
 \end{tikzpicture}}
 \subfigure{
  \begin{tikzpicture}
  \begin{axis}[
  width=4.0cm,
  height=3.4cm,
  ylabel={Precision},
  xlabel={},
  xmin=0, xmax=11,
  ymin=0.0975, ymax=0.1080,
  xtick={1,2,3,4,5,6,7,8,9,10},
  yticklabel style={/pgf/number format/.cd,fixed,precision=3},
  ytick={0.0980,0.1020,0.1060},
  ylabel style = {font=\scriptsize},
  xlabel style = {font=\scriptsize},
  tick label style={font=\scriptsize},
  scaled ticks=false,
  legend style={
  at={(1.05,0.5)}, 
  anchor=west,
  legend columns=1,
  font=\scriptsize,
  draw=none, 
  fill=none},
  title ={\textbf{LRURec}},
  title style={at={(-0.46,0.95)}, anchor=north west, font=\scriptsize},
  ]
 \addplot[color=8,
  solid,
  mark=*,
  mark options={solid},
  line width=1pt,
  mark size=1.2pt,
  smooth] coordinates { 
            (1, 0.1008)
            (2, 0.1013)
            (3, 0.1016)
            (4, 0.1011)
            (5, 0.1005)
            (6, 0.1007)
            (7, 0.1011)
            (8, 0.1008)
            (9, 0.1001)
            (10, 0.101)
  };
 \addplot[ color=77, 
  solid,
  mark=square,
  mark options={solid},
  line width=1pt,mark size=1.5pt,
  smooth] coordinates { 						
            (1, 0.1049)
            (2, 0.1044)
            (3, 0.1038)
            (4, 0.1047)
            (5, 0.1053)
            (6, 0.1055)
            (7, 0.1062)
            (8, 0.1068)
            (9, 0.1056)
            (10, 0.1054)
   };
 \addplot[ color=5, 
  solid,
  mark=diamond,
  mark options={solid},
  line width=1pt,mark size=1.5pt,
  smooth] coordinates { 						
            (1, 0.0985)
            (2, 0.1003)
            (3, 0.1)
            (4, 0.1005)
            (5, 0.0997)
            (6, 0.1001)
            (7, 0.1004)
            (8, 0.1002)
            (9, 0.0999)
            (10, 0.1)
        };
 \legend{AgentCF, DGDPO, KuaiSim}
 \end{axis}
 \end{tikzpicture}}
 \captionsetup{justification=centering}
 \caption{The performance of different user simulators in multi-round interactions on Books dataset.}
\label{fig:interaction_compare_books}
\end{figure*}
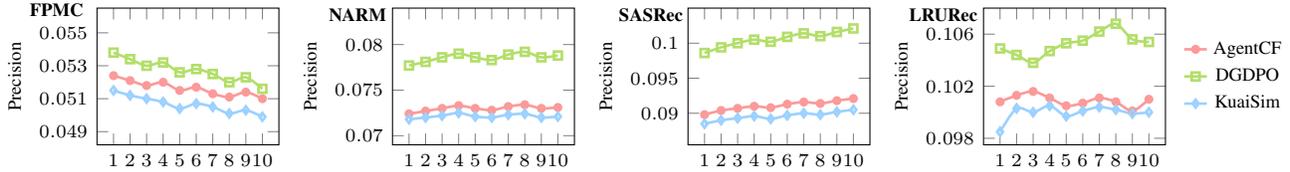

\subsubsection{Analysis of Multi-round Interactions (RQ2).}
Fig.~\ref{fig:interaction_analysis_book} depicts the performance of DGDPO across ten interaction rounds on Books dataset under different profile update strategies\footnote{Hereafter, we only show the results on Books due to space limitation, and similar trends can be noted on the other two datasets.}. 
The results unveils that dynamic profile update in multi-round interactions can further improve the simualtion accuracy. \textbf{(1)} For `\textit{w/o Update}', the performance generally exhibits minimal fluctuations across multi-round interactions and is the worst among all strategies. With profile unchanged,  `\textit{w/o Update}' fails to capture the evolution of user preference during multi-round interactions, thereby limiting the simulator's long-term fidelity. \textbf{(2)} `\textit{Update w/o GT}' exhibits performance divergence when interacting with different SRs. Specifically, with FPMC, the performance shows a declining trend. In contrast, with advanced deep learning-based SRs, we observe steady performance improvements. 
This occurs because advanced SRs rank ground truth items higher, and LLM-based user simulators, sensitive to item positions~\cite{hou2024large}, are more likely to interact with them when they are prominently placed, resulting in more accurate simulation. \textbf{(3)} `\textit{Update w/ GT}' performs the best among all methods. This is attributed to an idealized update strategy that always uses the ground truth item to update the user profile, regardless of the simulator's actions. 
However, in practical applications, ground truth items are often unavailable, which limits the applicability of this strategy. Therefore, it underscores the importance of developing more advanced user simulators that closely replicate real user behaviors, enabling more effective and reliable updates to the user profile during long-term interactions.

To further validate the effectiveness of our DGDPO in the multi-round interactions, we compare it against representative baselines. All simulators adopt the `\textit{w/o Update}' strategy and interact with four different SRs. Fig.~\ref{fig:interaction_compare_books} presents the results. \textbf{First}, DGDPO consistently achieves the highest Precision across every interaction round, indicating that it better captures user preferences and yields more realistic simulation. \textbf{Second}, although AgentCF underperforms DGDPO, it significantly outperforms traditional KuaiSim, reinforcing the effectiveness of LLM-based user simulators. 

\subsubsection{Ablation Study (RQ3).}
\begin{table}[t]
\footnotesize
\centering
\begin{tabular}{l|cccc}
\toprule
Variant & Prec & Recall & Acc & F1 \\ \midrule
\textit{w/o iteration} & 0.1702 & 0.7204 & 0.7262 & 0.2753 \\
\textit{w/o specialized LLM} & 0.1711 & 0.7219 & 0.7348 & 0.2760 \\
\textit{w/o fine-tuning} & 0.1803 & 0.7220 & 0.7404 & 0.2885 \\
\textit{w/o pre-training} & 0.2027 & 0.7411 & 0.7419 & 0.3183 \\
\midrule
DGDPO & \textbf{0.2058} & \textbf{0.7556} & \textbf{0.7570} & \textbf{0.3235} \\ \bottomrule
\end{tabular}
\caption{Ablation study of DGDPO on Books dataset.}
\label{tab:ablation-study}
\end{table}

Table~\ref{tab:ablation-study} presents the results of ablation study. \textbf{First}, DGDPO consistently outperforms its `\textit{w/o iteration}' variant, which conducts optimization only once without iteratively traversing interactions step-by-step for dynamic refinement. This showcases the importance of iterative optimization for progressively enhancing user profile quality and achieving superior fidelity. \textbf{Second}, we validate the effectiveness of our diagnostic module by comparing DGDPO with the `\textit{w/o specialized LLM}' variant, where the specialized LLM is replaced with a general-purpose LLM. The results confirm that domain-specific specialized LLM enables more reliable and precise diagnostic guidance. \textbf{Third}, by removing either domain-adaptive pre-training from the specialized LLM (`\textit{w/o pre-training}') or defect-specific fine-tuning (`\textit{w/o fine-tuning}'), the performance significantly drops compared to DGDPO, validating the usefulness of our designed training strategy. Notably, removing fine-tuning results in more substantial performance degradation, indicating a more critical role of defect-specific fine-tuning in profile defect identification. \textbf{Overall}, these findings confirm the effectiveness and necessity of all core components within our proposed DGDPO.

\subsubsection{Parameter Sensitivity Analysis (RQ4).}
\begin{table}[t]
\footnotesize
\centering
\begin{tabular}{c|cccc}
\toprule
Split Ratio \(\alpha\) & Prec & Recall & Acc& F1 \\ \midrule
0.5 & 0.2042 & 0.7522 & 0.7540 & 0.3212 \\
0.6 & \textbf{0.2058} & \textbf{0.7556} & \textbf{0.7570} & \textbf{0.3235} \\
0.7 & 0.1977 & 0.7253 & 0.7472 & 0.3107 \\
0.8 & 0.1902 & 0.7110 & 0.7411 & 0.3001 \\ \bottomrule
\end{tabular}
\caption{The impact of split ratio \(\alpha\) on Books dataset.}
\label{tab:parameter-analysis}
\end{table}

\begin{table}[t]
\centering
\footnotesize
\begin{tabular}{l|cccc}
\toprule
Specialized LLM & Prec & Recall & Acc & F1 \\ \midrule
Qwen3-0.6B (Ours) & \textbf{0.2058} & \textbf{0.7556} & \textbf{0.7570} & \textbf{0.3235} \\
Qwen3-1.7B & 0.2007 & 0.7541 & 0.7522 & 0.3170 \\ \bottomrule
\end{tabular}
\caption{The impact of model scale of the specialized LLM in the diagnostic module on Books dataset. }
\label{tab:foundation-model}
\end{table} 

We analyze the impact of split ratio \(\alpha\) on the performance of DGDPO. As shown in Table~\ref{tab:parameter-analysis}, the performance peaks when $\alpha=0.6$ and declines as the ratio deviates from this value. \textbf{(1)} A high \(\alpha\) value (e.g., 0.8) leaves limited data for the dynamic optimization, thereby restricting DGDPO’s capacity to detect and correct profile defects, which ultimately degrades the performance.
\textbf{(2)} 
A low \(\alpha\)  value (e.g., 0.5) results in profile initialization based on less data, increasing the risk of inaccuracy or incompleteness at the outset. Despite the availability of more data for subsequent optimization, the suboptimal starting point constrains final performance. This underscores the necessity of a balanced data allocation strategy to ensure both robust initialization and effective profile optimization.

We also examine the impact of the model scale of the specialized LLM in the diagnostic module.
Table~\ref{tab:foundation-model} compares the performance of our default Qwen3-0.6B model with a larger 1.7B variant. The results show that using the larger 1.7B model does not yield performance gains; instead, the Qwen3-0.6B model performs better across all metrics. A possible reason is that the profile defect identification is a specialized and fine-grained task. Thus, a smaller and more focused model like Qwen3-0.6B can be more effectively and efficiently fine-tuned using our designed training strategy. This finding indicates that for specialized components such as our diagnostic module, model scaling should be guided by task requirements rather than model size alone.

\subsubsection{Case Study (RQ5).}
\begin{figure}[t]
    \centering
    \includegraphics[width=1.0\linewidth]{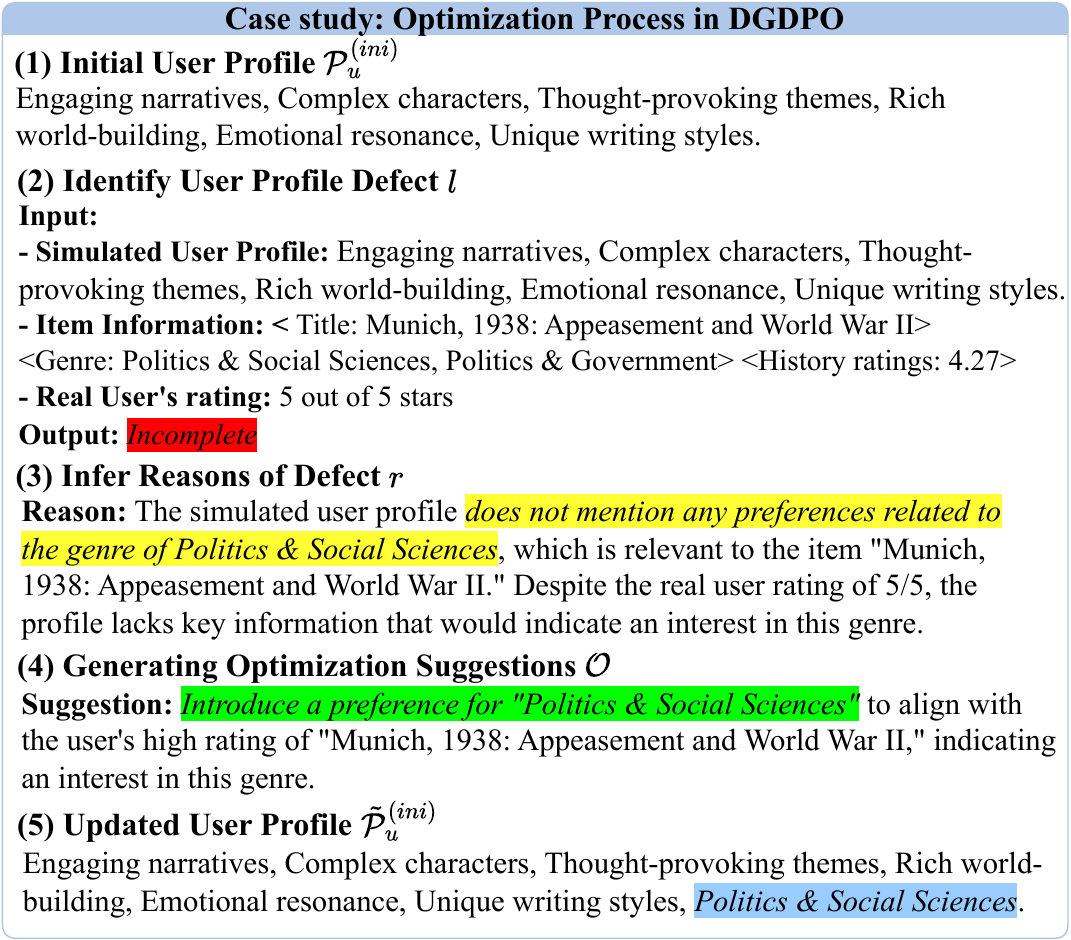}
    \caption{The case study on Books dataset.}
    \label{fig:case-study}
\end{figure}

Fig.~\ref{fig:case-study} illustrates the DGDPO optimization process with a case study from the Books dataset. A discrepancy (a highly-rated book being ignored by the simulator) triggers the diagnostic module to identify the profile as ``\textit{Incomplete}". The treatment module then infers the missing preference for ``Politics \& Social Sciences" and updates the profile accordingly.

\section{Conclusion}
In this paper, we propose DGDPO, a novel framework that constructs user profile through a dynamic optimization process to enhance simulation fidelity. DGDPO consists of two core modules: a specialized LLM-based diagnostic module calibrated through our novel training strategy first accurately identifies profile defects, which then guides a generalized LLM-based treatment module to generate targeted refinements. Moreover, we propose to incorporate DGDPO with sequential recommenders to enable realistic multi-round interactions, establishing a bidirectional evolution process between user profile and recommendation strategies. Experiments on three real-world datasets demonstrate the superiority of our DGDPO against the state-of-the-art methods.

\section{Acknowledgments}
This research is supported by the Australian Research Council (ARC) Discovery Project DP230100676 and partially supported by the Ministry of Education, Singapore, under its MOE AcRF Tier 1, SUTD Kickstarter Initiative (SKI 2021\_06\_12).

\bibliography{aaai2026}

@article{koren2009matrix,
  title={Matrix factorization techniques for recommender systems},
  author={Koren, Yehuda and Bell, Robert and Volinsky, Chris},
  journal={Computer},
  volume={42},
  number={8},
  pages={30--37},
  year={2009},
  publisher={IEEE}
}

@inproceedings{rendle2010factorizing,
  title={Factorizing personalized markov chains for next-basket recommendation},
  author={Rendle, Steffen and Freudenthaler, Christoph and Schmidt-Thieme, Lars},
  booktitle={Proceedings of the 19th international conference on World wide web (WWW)},
  pages={811--820},
  year={2010}
}

@article{hidasi2015session,
  title={Session-based Recommendations with Recurrent Neural Networks},
  author={Hidasi, B},
  journal={arXiv preprint arXiv:1511.06939},
  year={2015}
}

@article{harper2015movielens,
  title={The movielens datasets: History and context},
  author={Harper, F Maxwell and Konstan, Joseph A},
  journal={Acm transactions on interactive intelligent systems (TIIS)},
  volume={5},
  number={4},
  pages={1--19},
  year={2015},
  publisher={Acm New York, NY, USA}
}

@inproceedings{zhu2017next,
  title={What to do next: Modeling user behaviors by time-LSTM.},
  author={Zhu, Yu and Li, Hao and Liao, Yikang and Wang, Beidou and Guan, Ziyu and Liu, Haifeng and Cai, Deng},
  booktitle={Proceedings of the 26th International Joint Conference on Artificial Intelligence (IJCAI)},
  volume={17},
  pages={3602--3608},
  year={2017},
  organization={Melbourne, VIC}
}

@inproceedings{li2017neural,
  title={Neural attentive session-based recommendation},
  author={Li, Jing and Ren, Pengjie and Chen, Zhumin and Ren, Zhaochun and Lian, Tao and Ma, Jun},
  booktitle={Proceedings of the 26th ACM on Conference on Information and Knowledge Management (CIKM)},
  pages={1419--1428},
  year={2017}
}

@article{rohde2018recogym,
  title={Recogym: A reinforcement learning environment for the problem of product recommendation in online advertising},
  author={Rohde, David and Bonner, Stephen and Dunlop, Travis and Vasile, Flavian and Karatzoglou, Alexandros},
  journal={arXiv preprint arXiv:1808.00720},
  year={2018}
}

@inproceedings{tang2018personalized,
  title={Personalized top-n sequential recommendation via convolutional sequence embedding},
  author={Tang, Jiaxi and Wang, Ke},
  booktitle={Proceedings of the eleventh ACM international conference on web search and data mining (WSDM)},
  pages={565--573},
  year={2018}
}

@inproceedings{kang2018self,
  title={Self-attentive sequential recommendation},
  author={Kang, Wang-Cheng and McAuley, Julian},
  booktitle={2018 IEEE international conference on data mining (ICDM)},
  pages={197--206},
  year={2018},
  organization={IEEE}
}

@inproceedings{liu2018stamp,
  title={STAMP: short-term attention/memory priority model for session-based recommendation},
  author={Liu, Qiao and Zeng, Yifu and Mokhosi, Refuoe and Zhang, Haibin},
  booktitle={Proceedings of the 24th ACM SIGKDD international conference on knowledge discovery \& data mining (KDD)},
  pages={1831--1839},
  year={2018}
}

@article{ie2019recsim,
  title={Recsim: A configurable simulation platform for recommender systems},
  author={Ie, Eugene and Hsu, Chih-wei and Mladenov, Martin and Jain, Vihan and Narvekar, Sanmit and Wang, Jing and Wu, Rui and Boutilier, Craig},
  journal={arXiv preprint arXiv:1909.04847},
  year={2019}
}

@inproceedings{shi2019virtual,
  title={Virtual-taobao: Virtualizing real-world online retail environment for reinforcement learning},
  author={Shi, Jing-Cheng and Yu, Yang and Da, Qing and Chen, Shi-Yong and Zeng, An-Xiang},
  booktitle={Proceedings of the 33rd AAAI Conference on Artificial Intelligence (AAAI)},
  volume={33},
  number={01},
  pages={4902--4909},
  year={2019}
}

@inproceedings{chen2019generative,
  title={Generative adversarial user model for reinforcement learning based recommendation system},
  author={Chen, Xinshi and Li, Shuang and Li, Hui and Jiang, Shaohua and Qi, Yuan and Song, Le},
  booktitle={Proceedings of the 36th International Conference on Machine Learning (ICML)},
  pages={1052--1061},
  year={2019},
}

@inproceedings{shi2019pyrecgym,
  title={Pyrecgym: A reinforcement learning gym for recommender systems},
  author={Shi, Bichen and Ozsoy, Makbule Gulcin and Hurley, Neil and Smyth, Barry and Tragos, Elias Z and Geraci, James and Lawlor, Aonghus},
  booktitle={Proceedings of the 13th ACM Conference on Recommender Systems (RecSys)},
  pages={491--495},
  year={2019}
}

@inproceedings{sun2019bert4rec,
  title={BERT4Rec: Sequential recommendation with bidirectional encoder representations from transformer},
  author={Sun, Fei and Liu, Jun and Wu, Jian and Pei, Changhua and Lin, Xiao and Ou, Wenwu and Jiang, Peng},
  booktitle={Proceedings of the 28th ACM international conference on information and knowledge management (CIKM)},
  pages={1441--1450},
  year={2019}
}

@inproceedings{ni2019justifying,
  title={Justifying recommendations using distantly-labeled reviews and fine-grained aspects},
  author={Ni, Jianmo and Li, Jiacheng and McAuley, Julian},
  booktitle={Proceedings of the 2019 conference on empirical methods in natural language processing and the 9th international joint conference on natural language processing (EMNLP-IJCNLP)},
  pages={188--197},
  year={2019}
}

@inproceedings{zhu2019query,
  title={Query-based interactive recommendation by meta-path and adapted attention-gru},
  author={Zhu, Yu and Gong, Yu and Liu, Qingwen and Ma, Yingcai and Ou, Wenwu and Zhu, Junxiong and Wang, Beidou and Guan, Ziyu and Cai, Deng},
  booktitle={Proceedings of the 28th ACM International Conference on Information and Knowledge Management (CIKM)},
  pages={2585--2593},
  year={2019}
}

@inproceedings{huang2020keeping,
  title={Keeping dataset biases out of the simulation: A debiased simulator for reinforcement learning based recommender systems},
  author={Huang, Jin and Oosterhuis, Harrie and De Rijke, Maarten and Van Hoof, Herke},
  booktitle={Proceedings of the 14th ACM conference on recommender systems (RecSys)},
  pages={190--199},
  year={2020}
}

@inproceedings{zou2020pseudo,
  title={Pseudo Dyna-Q: A reinforcement learning framework for interactive recommendation},
  author={Zou, Lixin and Xia, Long and Du, Pan and Zhang, Zhuo and Bai, Ting and Liu, Weidong and Nie, Jian-Yun and Yin, Dawei},
  booktitle={Proceedings of the 13th International Conference on Web Search and Data Mining (WSDM)},
  pages={816--824},
  year={2020}
}

@inproceedings{zhao2021usersim,
  title={Usersim: User simulation via supervised generativeadversarial network},
  author={Zhao, Xiangyu and Xia, Long and Zou, Lixin and Liu, Hui and Yin, Dawei and Tang, Jiliang},
  booktitle={Proceedings of the 30th ACM Web Conference (TheWebConf)},
  pages={3582--3589},
  year={2021}
}

@inproceedings{luo2022mindsim,
  title={Mindsim: user simulator for news recommenders},
  author={Luo, Xufang and Liu, Zheng and Xiao, Shitao and Xie, Xing and Li, Dongsheng},
  booktitle={Proceedings of the 31st ACM Web Conference (TheWebConf)},
  pages={2067--2077},
  year={2022}
}

@article{wu2022gcrec,
  title={GCRec: Graph-augmented capsule network for next-item recommendation},
  author={Wu, Bin and He, Xiangnan and Zhang, Qi and Wang, Meng and Ye, Yangdong},
  journal={IEEE Transactions on Neural Networks and Learning Systems (TNNLS)},
  volume={34},
  number={12},
  pages={10164--10177},
  year={2022},
  publisher={IEEE}
}

@article{sun2022daisyrec,
  title={Daisyrec 2.0: Benchmarking recommendation for rigorous evaluation},
  author={Sun, Zhu and Fang, Hui and Yang, Jie and Qu, Xinghua and Liu, Hongyang and Yu, Di and Ong, Yew-Soon and Zhang, Jie},
  journal={IEEE Transactions on Pattern Analysis and Machine Intelligence (TPAMI)},
  volume={45},
  number={7},
  pages={8206--8226},
  year={2022},
  publisher={IEEE}
}

@inproceedings{afzali2023usersimcrs,
  title={UserSimCRS: a user simulation toolkit for evaluating conversational recommender systems},
  author={Afzali, Jafar and Drzewiecki, Aleksander Mark and Balog, Krisztian and Zhang, Shuo},
  booktitle={Proceedings of the 16th ACM International Conference on Web Search and Data Mining (WSDM)},
  pages={1160--1163},
  year={2023}
}

@article{park2023generative,
  title={Generative agents: Interactive simulacra of human behavior},
  author={Park, Joon Sung and O’Brien, Joseph C and Cai, Carrie J and Morris, Meredith Ringel and Liang, Percy and Bernstein, Michael S},
  journal={Proceedings of the 36th Annual ACM Symposium on User Interface Software and Technology (UIST)},
  year={2023}
}

@article{zhao2023kuaisim,
  title={KuaiSim: A comprehensive simulator for recommender systems},
  author={Zhao, Kesen and Liu, Shuchang and Cai, Qingpeng and Zhao, Xiangyu and Liu, Ziru and Zheng, Dong and Jiang, Peng and Gai, Kun},
  journal={Proceedings of the 37th Conference on Neural Information Processing Systems (NeurIPS)},
  volume={36},
  pages={44880--44897},
  year={2023}
}

@article{friedman2023leveraging,
  title={Leveraging large language models in conversational recommender systems},
  author={Friedman, Luke and Ahuja, Sameer and Allen, David and Tan, Zhenning and Sidahmed, Hakim and Long, Changbo and Xie, Jun and Schubiner, Gabriel and Patel, Ajay and Lara, Harsh and others},
  journal={arXiv preprint arXiv:2305.07961},
  year={2023}
}

@inproceedings{zhang2024generative,
  title={On generative agents in recommendation},
  author={Zhang, An and Chen, Yuxin and Sheng, Leheng and Wang, Xiang and Chua, Tat-Seng},
  booktitle={Proceedings of the 47th international ACM SIGIR conference on research and development in Information Retrieval (SIGIR)},
  pages={1807--1817},
  year={2024}
}

@inproceedings{sun2024large,
  title={Large language models for intent-driven session recommendations},
  author={Sun, Zhu and Liu, Hongyang and Qu, Xinghua and Feng, Kaidong and Wang, Yan and Ong, Yew Soon},
  booktitle={Proceedings of the 47th International ACM SIGIR Conference on Research and Development in Information Retrieval (SIGIR)},
  pages={324--334},
  year={2024}
}

@article{shu2024rah,
  title={RAH! RecSys--Assistant--Human: A Human-Centered Recommendation Framework With LLM Agents},
  author={Shu, Yubo and Zhang, Haonan and Gu, Hansu and Zhang, Peng and Lu, Tun and Li, Dongsheng and Gu, Ning},
  journal={IEEE Transactions on Computational Social Systems (TCSS)},
  year={2024},
  publisher={IEEE}
}

@inproceedings{zhang2024agentcf,
  title={Agentcf: Collaborative learning with autonomous language agents for recommender systems},
  author={Zhang, Junjie and Hou, Yupeng and Xie, Ruobing and Sun, Wenqi and McAuley, Julian and Zhao, Wayne Xin and Lin, Leyu and Wen, Ji-Rong},
  booktitle={Proceedings of the 33rd ACM on Web Conference (TheWebConf)},
  pages={3679--3689},
  year={2024}
}

@inproceedings{zhu2024reliable,
  title={How Reliable is Your Simulator? Analysis on the Limitations of Current LLM-based User Simulators for Conversational Recommendation},
  author={Zhu, Lixi and Huang, Xiaowen and Sang, Jitao},
  booktitle={Companion Proceedings of the ACM on Web Conference 2024 (WWW)},
  pages={1726--1732},
  year={2024}
}

@article{corecco2024llm,
  title={An LLM-based Recommender System Environment},
  author={Corecco, Nathan and Piatti, Giorgio and Lanzend{\"o}rfer, Luca A and Fan, Flint Xiaofeng and Wattenhofer, Roger},
  journal={arXiv preprint arXiv:2406.01631},
  year={2024}
}

@inproceedings{liu2024selfgnn,
  title={Selfgnn: Self-supervised graph neural networks for sequential recommendation},
  author={Liu, Yuxi and Xia, Lianghao and Huang, Chao},
  booktitle={Proceedings of the 47th International ACM SIGIR Conference on Research and Development in Information Retrieval (SIGIR)},
  pages={1609--1618},
  year={2024}
}

@inproceedings{yue2024linear,
  title={Linear recurrent units for sequential recommendation},
  author={Yue, Zhenrui and Wang, Yueqi and He, Zhankui and Zeng, Huimin and McAuley, Julian and Wang, Dong},
  booktitle={Proceedings of the 17th ACM International Conference on Web Search and Data Mining (WSDM)},
  pages={930--938},
  year={2024}
}

@inproceedings{hou2024large,
  title={Large language models are zero-shot rankers for recommender systems},
  author={Hou, Yupeng and Zhang, Junjie and Lin, Zihan and Lu, Hongyu and Xie, Ruobing and McAuley, Julian and Zhao, Wayne Xin},
  booktitle={Proceedings of the 46th European Conference on Information Retrieval (ECIR)},
  pages={364--381},
  year={2024},
  organization={Springer}
}

@inproceedings{li2024global,
  title={Global Heterogeneous Graph and Target Interest Denoising for Multi-behavior Sequential Recommendation},
  author={Li, Xuewei and Chen, Hongwei and Yu, Jian and Zhao, Mankun and Xu, Tianyi and Zhang, Wenbin and Yu, Mei},
  booktitle={Proceedings of the 17th ACM International Conference on Web Search and Data Mining (WSDM)},
  pages={387--395},
  year={2024}
}

@article{yu2024learning,
  title={Learning Neighbor User Intention on User--Item Interaction Graphs for Better Sequential Recommendation},
  author={Yu, Mei and Zhu, Kun and Zhao, Mankun and Yu, Jian and Xu, Tianyi and Jin, Di and Li, Xuewei and Yu, Ruiguo},
  journal={ACM Transactions on the Web (TWEB)},
  volume={18},
  number={2},
  pages={1--28},
  year={2024},
  publisher={ACM New York, NY}
}

@article{ebrat2024lusifer,
  title={Lusifer: LLM-based user simulated feedback environment for online recommender systems},
  author={Ebrat, Danial and Paradalis, Eli and Rueda, Luis},
  journal={arXiv preprint arXiv:2405.13362},
  year={2024}
}

@article{Balog:2024:FnTIR,
      author = {Krisztian Balog and ChengXiang Zhai},
    title = {User Simulation for Evaluating Information Access Systems},
 journal = {Foundations and Trends in Information Retrieval},
    year = {2024},
   volume = {18},
  doi = {10.1561/1500000098},
    url = {http://dx.doi.org/10.1561/1500000098},
    issn = {1554-0669},
      number = {1-2},
    pages = {1-261},
}

@article{wang2025user,
  title={User behavior simulation with large language model-based agents},
  author={Wang, Lei and Zhang, Jingsen and Yang, Hao and Chen, Zhi-Yuan and Tang, Jiakai and Zhang, Zeyu and Chen, Xu and Lin, Yankai and Sun, Hao and Song, Ruihua and others},
  journal={ACM Transactions on Information Systems (TOIS)},
  volume={43},
  number={2},
  pages={1--37},
  year={2025},
  publisher={ACM New York, NY}
}

@inproceedings{zhang2025llm,
  title={Llm-powered user simulator for recommender system},
  author={Zhang, Zijian and Liu, Shuchang and Liu, Ziru and Zhong, Rui and Cai, Qingpeng and Zhao, Xiangyu and Zhang, Chunxu and Liu, Qidong and Jiang, Peng},
  booktitle={Proceedings of the AAAI Conference on Artificial Intelligence (AAAI)},
  volume={39},
  number={12},
  pages={13339--13347},
  year={2025}
}

@inproceedings{cai2025agentic,
  title={Agentic feedback loop modeling improves recommendation and user simulation},
  author={Cai, Shihao and Zhang, Jizhi and Bao, Keqin and Gao, Chongming and Wang, Qifan and Feng, Fuli and He, Xiangnan},
  booktitle={Proceedings of the 48th International ACM SIGIR conference on Research and Development in Information Retrieval (SIGIR)},
  pages={2235--2244},
  year={2025}
}

@inproceedings{ma2025pub,
  title={PUB: An LLM-Enhanced Personality-Driven User Behaviour Simulator for Recommender System Evaluation},
  author={Ma, Chenglong and Xu, Ziqi and Ren, Yongli and Hettiachchi, Danula and Chan, Jeffrey},
  booktitle={Proceedings of the 48th International ACM SIGIR Conference on Research and Development in Information Retrieval (SIGIR)},
  pages={2690--2694},
  year={2025}
}

@article{wei2025mirroring,
  title={Mirroring Users: Towards Building Preference-aligned User Simulator with User Feedback in Recommendation},
  author={Wei, Tianjun and Guo, Huizhong and Du, Yingpeng and Sun, Zhu and Huang, Chen and Wang, Dongxia and Zhang, Jie},
  journal={arXiv preprint arXiv:2508.18142},
  year={2025}
}

@article{yang2025qwen3,
  title={Qwen3 technical report},
  author={Yang, An and Li, Anfeng and Yang, Baosong and Zhang, Beichen and Hui, Binyuan and Zheng, Bo and Yu, Bowen and Gao, Chang and Huang, Chengen and Lv, Chenxu and others},
  journal={arXiv preprint arXiv:2505.09388},
  year={2025}
}

@inproceedings{sun2025llm4rsr,
  title={LLM4RSR: Large Language Models as Data Correctors for Robust Sequential Recommendation},
  author={Sun, Yatong and Yang, Xiaochun and Sun, Zhu and Wang, Yan and Wang, Bin and Qu, Xinghua},
  booktitle={Proceedings of the AAAI Conference on Artificial Intelligence (AAAI)},
  volume={39},
  number={12},
  pages={12604--12612},
  year={2025}
}

@inproceedings{zhu2025llm,
  title={A llm-based controllable, scalable, human-involved user simulator framework for conversational recommender systems},
  author={Zhu, Lixi and Huang, Xiaowen and Sang, Jitao},
  booktitle={Proceedings of the ACM on Web Conference 2025 (WWW)},
  pages={4653--4661},
  year={2025}
}

\end{document}